\documentclass{ws-procs975x65}

\begin{document}

\title{THE RELATIVITY OF SPACE-TIME-PROPERTY}

\author{R. DELBOURGO$^*$}

\address{School of Mathematics and Physics, University of Tasmania,\\
Hobart, Tasmania 7001, Australia\\
$^*$E-mail: bob.delbourgo@utas.edu.au}

\begin{abstract}
We describe a geometrical way to unify gravity with the other natural forces by 
adding fermionic Lorentz scalar variables, characterising attribute or property, 
to space-time location. [With five such properties one can accommodate all 
known leptons and quarks.] Using just one property, viz. electricity, the general
relativity of such a scheme and its superscalar curvature automatically produces 
the Einstein-Maxwell Lagrangian and a cosmological term. By adding
more properties we envisage the geometrical unification of the standard
model with gravitation.
\end{abstract}

\keywords{Anticommuting coordinates; field properties; unified models.}

\bodymatter

\section*{Preamble}
It is a privilege and a delight to be able to celebrate with all of you here
Professor Freeman Dyson's life and achievements on the occasion of his
90th year. Little does Freeman know, but he has exerted a 
profound influence on my own career. Of course I was perfectly well aware of Dyson's
ground-breaking work on electrodynamics which was one of the first graduate courses
we were taught (by John C Taylor and Tom Kibble at Imperial College);  what really 
shaped my own career path was a chance remark made by  a physicist, namely
Hal Lewis, at Wisconsin University when I was a postdoc during 1963/64. The story 
goes that Freeman would bowl over at Summer 
Schools, such as the ones in Madison or Berkeley,  and ask what were the most significant 
problems in physics that year, in order of priority; he would then go through the list and 
spend his time solving them! It revealed a person who was unafraid to tackle the broadest 
range of subjects in his own inimitable way -- someone who had a panoramic view
of science and who was prepared to think  `outside the square', without getting
enmeshed in minutiae of one particular subarea. As if to reinforce that trait and
the breadth of his interests, Freeman visited us in Tasmania in April 1979 and 
gave  us a stimulating talk on how to mitigate carbon dioxide emissions by 
concerted planting of new trees; he would be well pleased that in Tasmania
today a good one third of the state is being preserved as native forest.

The work I am going to talk about was carried out with Mr Paul Stack, a brilliant 
PhD student with a great aptitude for physics, mathematics and computing: a rare
combination. The research has its genesis in the early days of supersymmetry 
(SUSY).  At that time I was tinkering with the idea of generalizing the auxiliary 
spinorial coordinate of SUSY to include 
internal symmetry labels (which now goes under the heading of $N$-extended
supersymmetry). Gell-Mann was interested in my attempts and, at a conference
in London, he had a look at this proposal. It did not take him long to say that the
idea would not fly as it would lead to spin state proliferation of an unacceptable
kind, and he was perfectly correct. So I put the idea on a back-burner and moved
on to other more fertile areas. Then, when I migrated to the backwoods of Tasmania,
a long way from the madding research crowd, I drew inspiration from Freeman 
Dyson to resist following the bandwagon -- a hopeless enterprise at my separation from 
the heartthrob of particle physics (internet notwithstanding) -- and try to do my own thing. 
Dyson's approach to science taught me not to keep up with the Jones'  but rather 
seek to be a Jones myself. Over the last few years I have tried resurrecting the idea of 
including particle attributes or properties mathematically with the aim of unifying gravity 
with the other natural forces., but in a way which differs very radically from spinorial 
SUSY. You can be the judge as to whether we have succeeded in this goal.

\section{Events}

A static universe is a contradiction in terms. Everything is inertial and
non-interacting, so we would not even be aware of its existence!
On the contrary, the universe evolves and its evolution is punctuated by
series of events, defined by
\begin{itemize}
\item WHERE and WHEN - location (x,y,z) and time (t)
\item WHAT - (ex)change of property ($\zeta = \xi+i\eta$)
\end{itemize}
My emphasis will be on the `what' and the overriding question is how to incorporate 
property or attribute mathematically. I shall
do so using anticommuting complex numbers, of which there are only a FINITE 
number, and by the same token the composition of these numbers remains finite.
Let us remind ourselves about such $a$-numbers..

Anticommuting operators were formulated in the 1920s but anticommuting
numbers, invented earlier by Grassmann for treating differential geometry, have been
used in physics only since about 1970, when BRST and SUSY came into prominence. 
{\em c-nos.} and {\em a-nos} are of course intimately connected with BE and FD statistics;
here is a little table which emphasizes their similarities and differences (I shall have 
more to say about the last entry presently).
\begin{center}
\begin{tabular}{|c|c|}
\hline
Bose-Einstein & Fermi-Dirac \\ \hline\hline
$xy = +yx$ & $\xi\eta = -\eta\xi$ \\ 
$[a,a^\dag] = 1$ & $\{a,a^\dag\} = 1$ \\ 
$(\mathrm{e}^{\alpha+\beta E}-1)^{-1}$ & $(\mathrm{e}%
^{\alpha+\beta E}+1)^{-1}$ \\ 
$\int \mathrm{e}^{cD^2c}dc\propto [\det D]^{-1}$ & 
$\int\mathrm{e}^{c^*Dc}d^2c\propto [\det D]^{+1}$ \\ 
O($2n)\sim$ Sp(-$2n$) & Sp($2n)\sim$ O(-$2n$) \\ 
\hline
\end{tabular}
\end{center}
Let me also remind you that BRST uses Lorentz scalar and vector a-numbers 
(pairs for ghosts and {$\overline{\mathrm{ghosts}}$}) but that SUSY uses Lorentz 
spinors. BRST is good for proving renormalizability and gauge invariance of gauge
models; there is no violation of the spin-statistics theorem for physical fields because 
asymptotic states are ghost-free. On the other side, SUSY also conforms to spin-statistics 
but uses {\em all} states in the asymptotic limit. Despite SUSY's great allure, its 
applicability has turned out to be problematic: nature (even the LHC) shows no signs of 
supersymmetric partner particles or states (photinos, squarks, gravitino,..) and this
has proved a great disappointment to me and to many others.

It is worth recapitulating the main theoretical attractions of SUSY:-
\begin{itemize}
\item consistent nontrivial higher group incorporating Poincare: 
$\{Q_{\alpha },Q_{\beta }\}=(\gamma .PC)_{\alpha \beta}$,
\item unifies bosons and fermions, 
\item cancellation of $\infty $s between boson and fermion loops, better
renormalizability,
\item allows supersymmetric generalization of gravity (SUGRA),
\item its extension allows `internal symmetries' to be incorporated via
generators $Q_{\alpha }^{n}$,
\item can be generalized to string/brane theory.
\end{itemize}
These are the reasons why so many physicists have persisted with investigating
SUSY, regardless and sometimes oblivious of experiment.

\section{Negative dimensions}
The most significant aspect of SUSY is that fermions and bosons act `oppositely'
to one another. Since $(\int\mathrm{e}^{cD^2c}dc)^{n_B}(\int\mathrm{e}^{\bar{\zeta}D
\zeta}d\zeta d\bar{\zeta})^{n_F} \propto D^{n_F-n_B}$, where $D$ is the Dirac operator,  
we can construe $c$-coordinates as adding to dimension and $a$-coordinates as 
subtracting dimension. This is also confirmed by group theory associated with $O(2n)$
and $Sp(2n)$ where an alteration in sign of $n$ allows one to continue Casimirs and
dimensions of certain classes of representation from one Lie algebra 
to the other \cite{n-n}.
\begin{quote}
``\emph{The Lord giveth and the Lord taketh away}''
\end{quote}
It does not really matter if the $a$-numbers are spinorial or scalar and I would like to
take advantage of that fact. By matching $c$-coordinates with $a$-coordinates (and 
likewise BE fields with FD fields) we get  may end up with {\em zero} nett dimensions 
(= \# dimensions of universe before the BIG BANG?).  
Therefore to the four $x^m$ of space-time I will add four $\zeta^\mu$ -- or equivalently 
add five $\zeta^\mu$, but only take half the states to get the statistics 
correct. I shall associate these Lorentz scalar $a$-nos. with `property' or `attribute',
leading to natural internal symmetries between these coordinates.

To obtain a sensible fundamental particle spectrum I have found it necessary to
make the following charge $Q$ and fermion number $F$ assignments:-
\[
Q(\zeta^{0,1,2,3,4}) = (0,1/3,1/3,1/3,-1);\, F(\zeta^{0,1,2,3,4}) =
(1,-1/3,-1/3,-1/3,1). 
\]
The $\zeta$ coordinates have been labelled from 0 to 4. Crudely we can regard
\begin{itemize}
\item label 0 as  `neutrinicity'
\item labels 1 -  3 as (antidown) `chromicity'
\item label 4 as `electricity'
\end{itemize}
Other properties are to be considered composites of these and it helps to regard
such attributes as the ingredients of a recipe. As you will recall, Rabi was heard to
say ``Who ordered that?'' when the muon was discovered. Well, in our scheme
the muon, tauon and more generally particle families are an integral part of the menu.

From these extended coordinates, superfields (functions of space-time and property) 
may be constructed \cite{SF}. Since the product of two {$a$-nos.} is a (nilpotent) {$c$-no.}, 
a Bose superfield {$\Phi$} should be a Taylor series in even powers of $\zeta,\bar{\zeta}$ 
and a Fermi superfield $\Psi_\alpha$ a series in odd powers of $\zeta,\bar{\zeta}$ 
--- up to the 5th: 
\[
\Phi(x,\zeta,\bar{\zeta})=\sum_{even\,r+\bar{r}}(\bar{\zeta})^
{\bar{r}} \phi_{(\bar{r}),(r)}(\zeta)^r; 
\]
\[
\Psi_\alpha(x,\zeta,\bar{\zeta})= \sum_{odd\, r+\bar{r}}(\bar
{\zeta})^{\bar{r}}\psi_{\alpha(\bar{r}),(r)}(\zeta)^r. 
\]
When forming actions as products of superfields we should integrate over all
spacetime as well as property to cover all possibilities and attribute changes.

The above expansions produce too many states, $\psi_\alpha$ and $\phi$, viz. 
256, so they need cutting down. (Had we only used four $\zeta$ we would not 
have been able to accommodate three generations as a matter of fact.)
Now a primary way to halve the number of states essentially is
to impose self-conjugation whereby 
\[
\psi_{(r),(\bar{r})} = \psi^{(c)}_{(\bar{r}),(r)}, 
\]
corresponding to reflection along the main diagonal in an $r,\bar{r}$ magic
square. Secondarily impose
(anti) self-duality corresponding to reflection about the cross diagonal,
specifically 
\[
\psi_{(r),(\bar{r})} = -\psi_{(r),(\bar{r})}^\times = - \psi_{(5-\bar{r}),(5-r)},
\]
\noindent noting that the dual ($\times$) of a field term has exactly the same 
$Q$ and $F$ as the field. For example, 
\[
(\bar{\zeta}^A\zeta_M\zeta_N)^{\times} = \frac{1}{3!}\epsilon_{JKLMN} \bar{
\zeta}^J\bar{\zeta}^K\bar{\zeta}^L.\frac{1}{4!}\epsilon^{ABCDE}\zeta_B
\zeta_C\zeta_D\zeta_E. 
\]
Apart from diminution of components through antiduality, we are able to exorcize
some unpleasant attribute combinations. Thus 
$\bar{\zeta}^0\bar{\zeta}^4\zeta_1\zeta_2\zeta_3$ and $\bar{
\zeta}^4\zeta_0\zeta_1\zeta_2\zeta_3$ are self-dual, so imposing antiduality
eliminates these nasties, since they have $F=3$ and $Q=-2$ respectively.

If we place these states in a kind of chessboard of dimensions $6\times 6$,
the fermions fall in the black squares and bosons in the white squares. The 
thing to note is that this scheme contains all known quarks and leptons, with
indications of a fourth generation. There are significant differences with the 
standard model however;-
\begin{itemize}
\item There are more than 3 leptons/D-quarks. These must necessarily be
heavy.
\item The first and second generations are electroweak doublets, but the
third generation is a triplet! Specifically ($i,j,k$ which run from one to three
are color labels below), 
\[
U_{1k}\sim \zeta^i\zeta^j\zeta^0,\qquad U_{2k}\sim
\zeta^i\zeta^j\zeta^0(\zeta^4\bar{\zeta}_4), \quad U_3^k\sim \zeta^k\bar{%
\zeta}_4\zeta^0 
\]
\[
D_{1k}\sim \zeta^i\zeta^j\zeta^4,\quad D_{2k}\sim
\zeta^i\zeta^j\zeta^4(\zeta^0\bar{\zeta}_0), \quad D_3^k\sim \zeta^k (\bar{%
\zeta}_0\zeta^0-\bar{\zeta}_4\zeta^4) 
\]
\[
\qquad\qquad\qquad\qquad\qquad\qquad\qquad X_3^k\sim \zeta^k \bar{\zeta}%
_0\zeta^4. 
\]
\item The CKM matrix will not be exactly unitary due to $X_3$ (charge -4/3); the
best place to search for an $X_3$ is presumably in electron-positron annihilation.
\end{itemize}

The mass matrix which affects quarks as well as leptons is due to a set of 
chargeless Higgs $\Phi$ field's expectation values; there are nine possibilities 
having $F=Q=0$ but associated with only one SU(5) invariant Yukawa coupling:-
\begin{itemize}
\item one $\phi_{(0)(0)} = \langle\phi\rangle$ ,
\item one $\phi_{(0)(4)} = \langle\phi_{1234}\rangle$ (standard
Higgs doublet expectation value),
\item three $\phi_{(1)(1)}=\langle\phi^0_0,\phi_4^4,\phi_i^i\rangle$
\item four $\phi_{(2)(2)}=\langle\phi_{04}^{04},\phi_{0k}^{0k},
\phi_{4k}^{4k},\phi_{ij}^{ij}\rangle,$
\end{itemize}
others being related by duality. This scheme is therefore more constrained
than the standard model. All this is by way of entree. 
I wish to offer the main course now.

\section{Graded General Relativity for Time-Space-Property}

You must be wondering where the gauge fields reside in such a scheme? It has 
probably already occurred to you that one may mimic SUSY and supergauge the 
massless free action for $\Psi $, but without added complication of spin. And indeed 
one can: in the time-honoured way, using the substitution rule for covariant derivatives 
whereby the generators of internal symmetry or shape-shifting property 
transformations are given by 
\[
T_A^B = \zeta_A\frac{\partial}{\partial\zeta_B} - \bar{\zeta}^B 
\frac{\partial}{\partial \bar{\zeta}^A} .
\]
But there is a more compelling way, which is fully geometrical and has the
benefit of incorporating gravity!

We will construct a fermionic version of Kaluza-Klein (KK) theory, this time without
needing to handle infinite modes which arise from compressed normal bosonic
coordinates. These are the significant points about the enlarged metric involving
the extended coordinate $X=(x,\zeta,\bar{\zeta})$:-
\begin{itemize}
\item One must introduce a fundamental length $\ell $ for the enlarged $X$,
because property $\zeta $ has no dimensions; this is tied to the gravity
scale $\kappa =\sqrt{8\pi G_{N}}$;
\item Gravity (plus gauge field products) fall within the $x-x$ sector,
gauge fields in $x-\zeta, x-\bar{\zeta}$ and the Higgs scalars must form a
matrix in $\zeta-\bar{\zeta}$;
\item The maximal gauge group is connected with the number of $\zeta$, 
so this is SU(5) in our scheme, ({\em  although nature seems only to gauge 
the standard subgroup for some unknown reason});
\item Gauge transformations are property rotations, dependent on space-time;
\item There is no place for a gravitino as spin is absent ($\zeta$ are
Lorentz scalar),;
\item There are necessarily a small {\em finite} number of modes.
\end{itemize}

To carry out this programme I need to introduce some basic notational niceties
first \cite{RDPJGT}. With the extended coordinate $X^M$, let $M=m$ (Roman) 
correspond to
space-time $x$ and let $M=\mu$ (Greek) correspond to property $\zeta, \bar{
\zeta}$. Set [$m$]=0 (no grading) and [$\mu$]=1 (grading). Thus $V^M A^N =
(-1)^{[M][N]} A^N V^M$.
Now we can revisit general relativity \emph{taking great care with ordering and
sign factors}! Our rule is always to take left derivatives, like $\frac{\partial}{\partial
X^M}A_N$ but we have to reconcile this with traditional GR notation, $A_{N,M}$,
which is ingrained but \emph{back to front}! This introduces sign factors and we
have to live with that.

Transformation laws for vectors $V$ \& tensors $T$ (such as the metric $G$%
), etc. read 
\[
V^{\prime M}=V^N\frac{\partial X^{\prime M}}{\partial X^N},\, V_M^{\prime} = 
\frac{\partial X^N}{\partial X^{\prime M}}V_N,\, \mathrm{~so~} V^NV_N 
\mathrm{~is~invariant}, 
\]
\[
T^{\prime}_{MN}=(-1)^{[R]([S]+[N])}\frac{\partial X^R}{\partial X^{\prime M}}
\frac{\partial X^S}{\partial X^{\prime N}}T_{RS},\,\, \mathrm{~~so~}
ds^2=dX^N dX^MG_{MN} \mathrm{~is~invariant}. 
\]
We can use the metric tensor and its inverse to raise and lower indices: 
\[
V^MG_{MN}\equiv V_N,\quad G^{MN}V_N=V^M,\quad {T_M}^N =
(-1)^{[M]([N]+[L])}G^{NL}T_{ML}, 
\]
with 
\[
G^{MN}G_{NL}={\delta^M}_L= (-1)^{[M]}{\delta_L}^M,\qquad
(-1)^{[N]}G_{MN}G^{NL} = {\delta_M}^L 
\]
and noting that $G$ and its inverse are graded symmetric: 
\[
G_{MN}=(-1)^{[M][N]}G_{NM}, \quad G^{LM}=(-1)^{[L][M]}G^{ML}. 
\]

\section{Covariant derivatives and the Riemann supertensor}

With these conventions one can establish the rules for
covariant differentiation: 
\[
A_{M;N} = (-1)^{[M][N]}A_{M,N} - \Gamma_{MN}{}^L A_L, 
\]
\[
{A^M}_{;N} = (-1)^{[M][N]}[{A^M}_{,N} + A^L{\Gamma_{LN}}^M], 
\]
\[
{T_{LM}}_{;N}=(-1)^{[N]([L]+[M])}\big[T_{LM,N}-{\Gamma_{NL}}^K T_{KM} 
-(-1)^{[L]([M]+[K])}{\Gamma_{NM}}^KT_{LK}\big],
\]
etc. where the {\em Christoffel connection} is given by 
\[
\Gamma_{MN}{}^K = \big[(-1)^{[M][N]}G_{ML,N}  + G_{NL,M}-
(-1)^{[L]([M]+[N])} G_{MN,L} \big] (-1)^{[L]}G^{LK}/2. 
\]
From this may be derived the generalized Riemann tensor $\cal R$: 
\[
(-1)^{[J]}A_J {\cal R}^J{}_{KLM} = A_{K;L;M} - (-1)^{[L][M]} A_{K;M;L} 
\]
whereupon $\cal R$ can be expressed in terms of the connections. 
\[
{\cal R}^J{}_{KLM}=(-1)^{[J]([K]+[L]+[M])}\big[(-1)^{[K][L]}
\Gamma_{KM}^J{}_{,L}- (-1)^{([K]+[L])[M]}\Gamma_{KL}^J{}_{,M} 
\]
\[
\qquad\qquad\qquad\qquad\qquad\qquad + (-1)^{[L][M]}
\Gamma_{KM}{}^R\Gamma_{RL}{}^J-\Gamma{}_{KL}{}^R\Gamma_{RM}{}^J \big].
\]
As a further check, we may derive the lowered tensor 
\[
\mathcal{R}_{JKLM} =
(-1)^{([I]+[J])([K]+[L]+[M])}\mathcal{R}^I{}_{KLM}G_{IJ},
\]
and check its symmetry properties, 
\begin{eqnarray*}
\mathcal{R}_{KJLM} &=& - (-1)^{[J][K]} \mathcal{R}_{JKLM} , \\
\mathcal{R}_{JKML} &=& - (-1)^{[L][M]} \mathcal{R}_{JKLM}, \\
\mathcal{R}_{LMJK} &=& (-1)^{([J]+[K])([L]+[M])} \mathcal{R}_{JKLM}.
\end{eqnarray*}
as well as the cyclicity property (\emph{first Bianchi identity}): 
\[
(-1)^{KM} \mathcal{R}_{JKLM} + (-1)^{ML} \mathcal{R}_{JMKL} + (-1)^{LK} 
\mathcal{R}_{JLMK} = 0. 
\]
The \emph{second (differential) Bianchi identity} can also be established, 
\[
(-1)^{[L][N]}\mathcal{R}_{JKLM;N} +(-1)^{[N][M]}\mathcal{R}_{JKNL;M} 
+(-1)^{[M][L]}\mathcal{R}_{JKMN;L} = 0. 
\]
The \emph{Ricci} tensor is arrived at via the contraction: 
\[
\mathcal{R}_{KM}= (-1)^{[K][L]} G^{LJ}\mathcal{R}_{JKLM},\mathrm{~with~} 
\mathcal{R}_{KM}=(-1)^{[K][M]}R_{MK}, 
\]
and the full superscalar curvature is obtained as $\mathcal{R}\equiv G^{MK}%
\mathcal{R}_{KM}$. Finally to get the \emph{Einstein} tensor $\mathcal{G}$
and its vanishing covariant divergence, contract out the second Bianchi
identity. One finds automatically: 
\[
\mathcal{R}_{;N} = 2 (-1)^{[M][N]} \mathcal{R}^{M}{}_{N;M}. 
\]
or $\mathcal{G}^{M}{}_{N;M}=0$ where 
\[
{\mathcal{G}^M}_N = {\mathcal{R}^M}_N -{\delta^M}_N\mathcal{R}/2. 
\]

\section{Frame vectors and metric}

Now let me focus on just one property, namely {\em electricity}, so there
is only one $\zeta$ and its conjugate $\bar{\zeta}$ and we need not bother with
indices on $\zeta$. In flat space 
\[
ds^2 = dX^AdX^B\eta_{BA} = dx^a dx^b\eta_{ba} +  d\zeta d\bar{\zeta} \eta_{%
\bar{\zeta}\zeta} +d\bar{\zeta}d\zeta \eta_{\zeta\bar{\zeta}},  
\]
where $\eta_{\zeta\bar{\zeta}}=-\eta_{\bar{\zeta}\zeta}=\ell^2/2$ and $%
\eta_{ba}$ is Minkowskian. To curve the space, let us be guided by Kaluza-Klein 
and introduce frame vectors $\mathcal{E}$, allowing for property curvature
coefficients $c_i$: 
\[
G_{MN} = (-1)^{[B]+[B][N]}{\mathcal{E}_M}^A \eta_{A B} {\mathcal{E}_N}^B, 
\]
with 
\begin{eqnarray*}
{\mathcal{E}_m}^a\!\!\!&=&\!\!\!(1\!\!+\!c_1\bar{\zeta}\zeta)
{e_m}^a},\,\,\, {\mathcal{E}_m}^\zeta\!\!=\!\!-{ie\bar{\zeta}A_m
,\,\,\, {\mathcal{E}_m}^{\bar{\zeta}}\!\!=\! -ieA_m\zeta, 
\nonumber \\
{\mathcal{E}_\zeta}^a &=& 0, \quad{\mathcal{E}_\zeta}^\zeta=0,\quad {%
\mathcal{E}_\zeta}^{\bar{\zeta}}={(1+c_2\bar{\zeta}\zeta)}, 
\nonumber \\
{\mathcal{E}_{\bar{\zeta}}}^a &=&0, \quad {\mathcal{E}_{\bar{\zeta}}}^\zeta
= -{(1+c_2\bar{\zeta}\zeta)},\quad {\mathcal{E}_{\bar{\zeta}}}^{%
\bar{\zeta}} = 0,
\end{eqnarray*}

Putting this all together results in the following metric: 
\[
G_{MN} = \left( 
\begin{array}{ccc}
{G_{m n}} & {G_{m\zeta}} & {G_{m\bar\zeta}}
\\ 
{G_{\zeta n}} & 0 & {G_{\zeta \bar\zeta}} \\ 
{G_{\bar\zeta n}} & {G_{\bar\zeta \zeta}} & 0%
\end{array}
\right)\qquad \mathrm{where}
\]
\begin{eqnarray*}
{G_{mn}} &=& g_{mn}(1 + 2 c_1 \bar{\zeta} \zeta)+ e^2\ell^2 A_m
A_n \bar{\zeta}\zeta,  \nonumber \\
{G_{m\zeta}} &=& {G_{\zeta m}}= -ie\ell^2 A_m \bar{%
\zeta}/2,  \nonumber \\
{G_{m\bar\zeta}}&=&{G_{\bar\zeta m}}=-ie\ell ^2
A_m\zeta/2,  \nonumber \\
{G_{\zeta\bar\zeta}} &=& -{G_{\bar\zeta \zeta}} =
\ell^2(1+2 c_2 \bar{\zeta} \zeta)/2).
\end{eqnarray*}

A few pertinent observations about the covariant metric $G$ are appropriate.
\begin{itemize}
\item the charge coupling $e$ accompanies the e.m. potential $A$;
\item the constants $c_i$ provide phase invariant property curvature -- like
mass in the Schwarzschild metric. They may be expectation of Higgs fields
or possibly be associated with dilatons;
\item there is an intrinsic length scale, which will be tied to gravity; 
\item there are \emph{no gravitinos}; such fields carry a spinor index, but
$\zeta$ are \emph{scalar} so any gravitinos would spoil Lorentz invariance
and cannot appear in the metric;
\item one might consider including a term like $C_m(1 + \alpha\bar{\zeta}\zeta)$
to the $x-\zeta$ frame vector where $C_m$ is an anticommuting vector, as one
meets in \emph{quantized} gravity (Feynman's vector ghosts). Similarly we could
add $\bar{C}\zeta+\bar{\zeta}C$ where $C$ are scalar ghosts in the purely 
property sector, connecting with BRST for gauge vector quantization. We have not
done so, since we are confining ourselves to classical gravity/e.m.
\item otherwise the metric is \emph{as general as it can be};
\item {\em fermions are distinct -- they carry a spinor index--\&must be treated 
separately}.
\end{itemize}

We can similarly determine the elements of the inverse metric: 
\begin{eqnarray*}
{G^{mn}}&=& g{}^{m}{}^{n}(1 - 2 c_1\bar{\zeta}\zeta),  \nonumber
\\
{G^{m \zeta}}&=&{G^{\zeta m}} = i eA{}^{m} \zeta, 
\nonumber \\
{G^{m\bar\zeta}}&=&{G^{\bar\zeta m}}=-ieA{}^{m}\bar{%
\zeta},  \nonumber \\
{G^{\zeta\bar\zeta}}&=&\!\!-{G^{\bar\zeta \zeta}}=
2(1-2 c_2\bar{\zeta}\zeta)/\ell^2\!-\! e^2A^mA_m \bar{\zeta}\zeta,
\end{eqnarray*}
and go on to determine the Christoffel connections. I shall not bother to list these;
some are contained in earlier papers \cite{AL}.

Before we can get the action we will also need the superdeterminant of the
metric: 
\[
s\det(G_{MN}) = \frac{4}{\ell^4} \det(g{}_{m}{}_{n}) \left[1 + (8 c_1-4c_2)%
\bar{\zeta} \zeta\right]
\]
or for short, 
\[
\sqrt{-G..} = \frac{2}{\ell^2} \sqrt{-g..} \left[1 + (4c_1-2c_2) \bar{\zeta}
\zeta\right].
\]
The absence of the gauge potential should be noted (because $\mathcal{E}$ is
triangular).

In connection with the super-determinant, one can establish  that 
$(\sqrt{G..})_{,M}=\sqrt{G..}(-1)^{[N]}{\Gamma_{MN}}^N$ and, from the
definition of the Christoffel symbol, show that the scalar curvature can be reduced to
the analogue of the Palatini form:
\begin{equation*}
\sqrt{G..}{\cal R} \rightarrow (-1)^{[L]}\sqrt{G..}G^{MK}\left[ (-1)^{[L][M]}
  {\Gamma_{KL}}^N{\Gamma_{NM}}^L-{\Gamma_{KM}}^N{\Gamma_{NL}}^L \right].
\end{equation*}

\section{Gauge changes as property transformations}

Make a spacetime dependent U(1) phase transformation in the property sector: 
\[
x^\prime = x; \;\;\;  \zeta^\prime = e^{i \theta(x)} \zeta; \;\;\;
\bar\zeta^\prime = e^{-i \theta(x)} \bar\zeta. 
\]
From the general transformation rules of { $G_{m\zeta}$} we find 
\[
eA^\prime{}_m = eA_m + \partial_m \theta, 
\]
showing the field $A_m$ acts as a gauge field under variations in electric
phase. [This can be checked for all components of the metric $G_{MN}$.] But
$G^{mn}$ remains unaffected and thus is gauge-invariant.

The same comments apply to $\mathcal{R}_{mn}$ and $\mathcal{R}^{mn}$; the
former varies with gauge but the latter does not.

\section{The Ricci tensor and gravitational--e.m. action}

From $G$ and the evaluation of $\Gamma$ we can determine the
full Riemann curvature supertensor $\mathcal{R}_{JKLM}$ and the Ricci tensor 
$\mathcal{R}_{KM}= (-1)^{[K][L]} G^{LJ}\mathcal{R}_{JKLM}$; whence the
superscalar $\mathcal{R}=G^{MK}\mathcal{R}_{KM}$. It is all a matter of
cranking the handle, but an algebraic computer program which works out
these quantities with no errors is of enormous assistance and Paul Stack has
been instrumental in developing such a program using {\em Mathematica}.

To see how electromagnetism emerges geometrically ignore spacetime curvature
initially but not property curvature. The spacetime component of
contravariant Ricci reduces to the gauge independent pair: 
\[
\mathcal{R}^{mn}\!\!=4g^{mn}c_1[1+(2c_2-6c_1)\bar{\zeta}\zeta]/\ell^2
\!\!-e^2 \ell^2 F^{ml}{F^n}_l\, \bar{\zeta}\zeta/2, 
\]
and the curvature superscalar collapses to 
\[
\mathcal{R}= 8[4c_1\!-\!3c_2\!+\!c_1(8c_2\!-\!10c_1)\,\bar{\zeta}\zeta]
/\ell^2 - e^2 \ell^2 F^{nl}F_{nl}\bar{\zeta}\zeta/4. 
\]

Including the super-determinant the total Lagrangian density for
electromagnetic property emerges: 
\[
\mathcal{L}=\int \!\!\sqrt{-G..}d\zeta d\bar{\zeta}\,\mathcal{R}\propto -%
\frac{1}{4}F_{mn}F^{mn}+\frac{48(c_{1}-c_{2})^{2}}{e^{2}\ell ^{4}},
\]
and the Einstein tensor in flat spacetime reduces to: 
\[
\int \!d\zeta d\bar{\zeta}\sqrt{-G..}(\mathcal{R}^{km}-\mathcal{R}%
\;G^{km}/2)\propto 
\]
\[
\left[ 48c_{2}(c_{1}\!-\!c_{2})g^{km}/e^{2}\ell ^{4}\!-\!(F^{kl}
{F^m}_l - F_{ln}F^{ln}g^{km}/4)\right] .
\]
The familiar expression for the e.m. stress tensor, $T^{km}\equiv 
{F^k}_l F^{lm}+F_{nl}F^{nl}g^{km}/4$, {\em becomes part of the geometry}.

Now include gravity by curving spacetime ($\eta_{mn}\rightarrow g_{mn}(x)$).
\[
\mathcal{L}=\int\! \sqrt{-G..}d \zeta d \bar\zeta\,\mathcal{R}  =2e^2\sqrt{%
-g..} \left[  \frac{2(c_1-c_2)R}{e^2\ell^2} -\frac{F_{mn}F^{mn}}{4} +  \frac{%
48(c_1-c_2)^2}{e^2\ell^4}\right]. 
\]
Spot the Newtonian constant and the cosmological term, 
\[
16\pi G_N\equiv \kappa^2 = e^2\ell^2/2(c_1-c_2), \quad \Lambda =
12(c_2-c_1)/\ell^2.
\]
The field equations arise from varying $G$ but, since $\delta G_{MN}
= \delta g_{mn}(1+2c_1\bar{\zeta}\zeta)$, we must evaluate the following
integral in order to obtain the gravitational equation:
\begin{eqnarray*} \label{einsteintensor}
0=\int d\zeta d\bar{\zeta} \sqrt{-G..}&(&1+2c_1\bar{\zeta}\zeta)({\cal R}^{km} 
-G^{km}{\cal R} / 2 )\\
&=&\sqrt{-g..}\left[\frac{4(c_1-c_2)}{\ell^2}(R^{km}\!\!-\frac{1}{2}g^{km}R)
 -  T^{km}\! - \frac{48(c_1\!-\!c_2)^2}{\ell^4}g^{km}\!\right].
\end{eqnarray*}
This is just what we would have obtained from $\cal L$. In any case we see
that the universal coupling of gravity to stress tensors $T$ has a factor
$8\pi G_N\equiv \kappa^2/2 = e^2\ell^2/4(c_1-c_2) > 0$.  The result is
to make the cosmological term go negative and, what is probably worse,
it has a value which is inordinately larger than the tiny experimental 
value found by analyses of supernovae! (All cosmological terms derived 
from particle physics, except for exactly zero, share the same problem). 
Numerically speaking, $\kappa\simeq 5.8\times 10^{-19}$ (GeV)$^{-1}$ 
means $\ell \sim 10^{-18}$ (GeV)$^{-1}$ is Planckian in scale. Of course
the magnitude of the miniscule cosmological constant $\Lambda\sim4\times
10^{-84}$ (GeV)$^2$  is at variance with Planckian expectations by the usual 
factor of $10^{-120}$, which is probably the most mysterious natural ratio. So 
far as our scheme is concerned, we are disappointed but not particularly 
troubled by the wrong sign of $\Lambda$ because it can readily be 
reversed by  extra property curvature coefficients when we enlarge the 
number of properties (as we have checked when enlarging the 
number of properties to at least two). The magnitude of $\Lambda$ is 
quite another matter because it will require some extraordinary fine-tuning, 
even after fixing the sign.

Note added:  since giving the talk, Stack and I have extended our work to two 
properties and derived the SU(2) Yang-Mills Lagrangian, united geometrically 
to gravity. A few more steps will take us to electroweak theory, so stay tuned.

\section*{Acknowledgements}
I am most grateful to the organizers for the invitation to attend this memorable
conference and for the opportunity to deliver this lecture, outlining our progress
in this area. I would also like to acknowledge Paul Stack's very substantial
contributions to these developments (especially his marvellous computer
program for carrying out the algebraic manipulations) and Peter Jarvis' 
encouragement and perceptive comments.

\end{document}